\pdfoutput=1

\documentclass[aps,pra,10pt,reprint,amsmath,superscriptaddress]{revtex4-2}
\usepackage{graphicx}
\usepackage{amsmath}
\usepackage[separate-uncertainty=true,multi-part-units=single,range-phrase=--]{siunitx}
\DeclareSIUnit\clight{\text {\ensuremath {c}}}
\DeclareSIUnit\eVperc{\eV\per\clight}
\usepackage{hyperref}
\usepackage[nameinlink,capitalize]{cleveref}

\begin{document}
\title{Development of an optimal laser for chirp cooling of positronium based on chirped pulse-train generator}
\date{\today}
\author{Kenji Shu}
\affiliation{Photon Science Center, Graduate School of Engineering, The University of Tokyo, 2-11-16 Yayoi, Bunkyo-ku, Tokyo 113-0032, Japan}
\affiliation{Department of Applied Physics, School of Engineering, The University of Tokyo, 7-3-1 Hongo, Bunkyo-ku, Tokyo 113-8656, Japan}
\affiliation{Department of Applied Physics, Faculty of Engineering, The University of Tokyo, 7-3-1 Hongo, Bunkyo-ku, Tokyo 113-8656, Japan}
\author{Naoki Miyamoto}
\affiliation{Department of Applied Physics, School of Engineering, The University of Tokyo, 7-3-1 Hongo, Bunkyo-ku, Tokyo 113-8656, Japan}
\author{Yuto Motohashi}
\affiliation{Department of Applied Physics, Faculty of Engineering, The University of Tokyo, 7-3-1 Hongo, Bunkyo-ku, Tokyo 113-8656, Japan}
\author{Ryosuke Uozumi}
\affiliation{Department of Applied Physics, School of Engineering, The University of Tokyo, 7-3-1 Hongo, Bunkyo-ku, Tokyo 113-8656, Japan}
\author{Yohei Tajima}
\affiliation{Department of Applied Physics, School of Engineering, The University of Tokyo, 7-3-1 Hongo, Bunkyo-ku, Tokyo 113-8656, Japan}
\author{Kosuke Yoshioka}
\email{yoshioka@fs.t.u-tokyo.ac.jp}
\affiliation{Photon Science Center, Graduate School of Engineering, The University of Tokyo, 2-11-16 Yayoi, Bunkyo-ku, Tokyo 113-0032, Japan}
\affiliation{Department of Applied Physics, School of Engineering, The University of Tokyo, 7-3-1 Hongo, Bunkyo-ku, Tokyo 113-8656, Japan}
\affiliation{Department of Applied Physics, Faculty of Engineering, The University of Tokyo, 7-3-1 Hongo, Bunkyo-ku, Tokyo 113-8656, Japan}

\begin{abstract}
  We report the development and characterization of a pulsed \SI{243}{nm} laser that is optimal for the cooling of positronium~(Ps).
  The laser, which is based on the recent chirped pulse-train generator (CPTG) demonstrated by K. Yamada \textit{et al.} (Phys. Rev. Appl. \textbf{16}, 014009 (2021)), was designed to output a train of pulses with linewidths of \SI{10}{GHz}, and with the center frequency of each pulse shifting upward~(up-chirped) in time by \SI{4.9e2}{\GHz \per \us}.
  These parameters were determined by the mechanism of chirp cooling, which is the best scheme for cooling many Ps atoms to the recoil temperature of laser cooling. 
  To achieve the designed performance, we drove an optical phase modulator in the CPTG with a deep modulation depth based on the operating principle of the cooling laser.
  Time-resolved spectroscopic measurements confirmed that the developed laser satisfied the chirp rate and linewidth requirements for efficient chirp cooling.
  Combined with pulse energy of hundreds of microjoules, we believe that the experimental demonstration of Ps laser cooling has become possible using realistic methods for the generation and velocity measurement of Ps.
\end{abstract}

\maketitle

\section{Introduction}
Positronium~(Ps), which is the bound state of an electron and a positron, is a unique system for precise investigation of fundamental physics~\cite{karshenboim_precision_2005,adkins_precision_2022}.
Its purely leptonic composition allows precise calculations of eigenenergies and lifetimes.
Comparisons between the calculated energy intervals and their precise measurements can be stringent tests of quantum electrodynamics and probes for new physics, such as the asymmetry between matter and antimatter.
For most intervals between the energy levels of Ps, the measurement precision is currently inferior to that of calculations~\cite{adkins_precision_2022, cassidy_experimental_2018,fee_measurement_1993,ishida_new_2014,gurung_precision_2020}.
One of the difficulties in these spectroscopic measurements arises from the light mass of Ps.
As Ps is the lightest atom, its velocity distribution is very broad at typical temperatures, which dominantly contributes to measurement errors in both the optical and microwave regions~\cite{fee_measurement_1993,gurung_precision_2020}.
Cooling of Ps will therefore be a breakthrough achievement in increasing the measurement precision and even in achieving Bose--Einstein condensation combined with a dense cloud of Ps~\cite{platzman_possibilities_1994,morandi_bose-einstein_2014,shu_study_2016}.
The conventional technique uses thermalization processes between energetic Ps atoms and cryogenic materials~\cite{shu_observation_2021,guatieri_classical_2022}.
The lowest temperature of the Ps atom clouds emitted into vacuum is currently limited to approximately \SI{100}{K}~\cite{mariazzi_positronium_2010,shu_observation_2021,guatieri_time--flight_2021}.

Laser cooling is a promising technique for cooling Ps to temperatures significantly lower than the current limit~\cite{liang_laser_1988,kumita_study_2002,zimmer_positronium_2021}.
One advantage is the large recoil velocity achieved by absorbing the momentum of a photon because of its light mass.
This is particularly important for Ps, which is an unstable atom with a decay lifetime of \SI{142}{ns} in the long-lived $1^3\mathrm{S}_1$ state~\cite{adkins_order_2000,kataoka_first_2009}.
Rapid cooling of initially hot Ps is considered possible using the 1S--2P transition in a time duration comparable to its lifetime.
However, the light mass makes it difficult to prepare an appropriate laser.
A laser at a wavelength of \SI{243}{nm} with both a broadband spectrum and long duration is required.
The difficulty in building such a laser is one of the reasons why the laser cooling of Ps has never been performed.
One of the recent attempts was the development of the chirped pulse-train generator (CPTG)~\cite{yamada_theoretical_2021}.
The CPTG had the desired features for cooling Ps, especially by utilizing a time-dependent shift of the laser spectrum to improve the cooling efficiency~\cite{shu_study_2016}.
However, the developed CPTG could only decelerate a limited portion of components in a broad Doppler profile.
Further improvements are necessary to cool the majority of Ps atoms, which are widely Doppler broadened, down to the fundamental limit of laser cooling.
In this study, we report the development of a laser optimized for cooling Ps.
The design is based on the chirp-cooling scheme, which is the most suitable for cooling many Ps atoms down to the recoil limit temperature of laser cooling, as shown below.
A key point of the development was the enhancement of the spectral broadening of the CPTG by an order of magnitude.
The enhancement was achieved by a pulsed deep drive of an optical phase modulator.
We also improved the methodology for characterizing the structure of the developed laser in the time and spectral domains to directly confirm that the laser had the desired properties for chirp cooling of Ps.

\begin{figure*}
  \centering
  \includegraphics[width=0.9\linewidth]{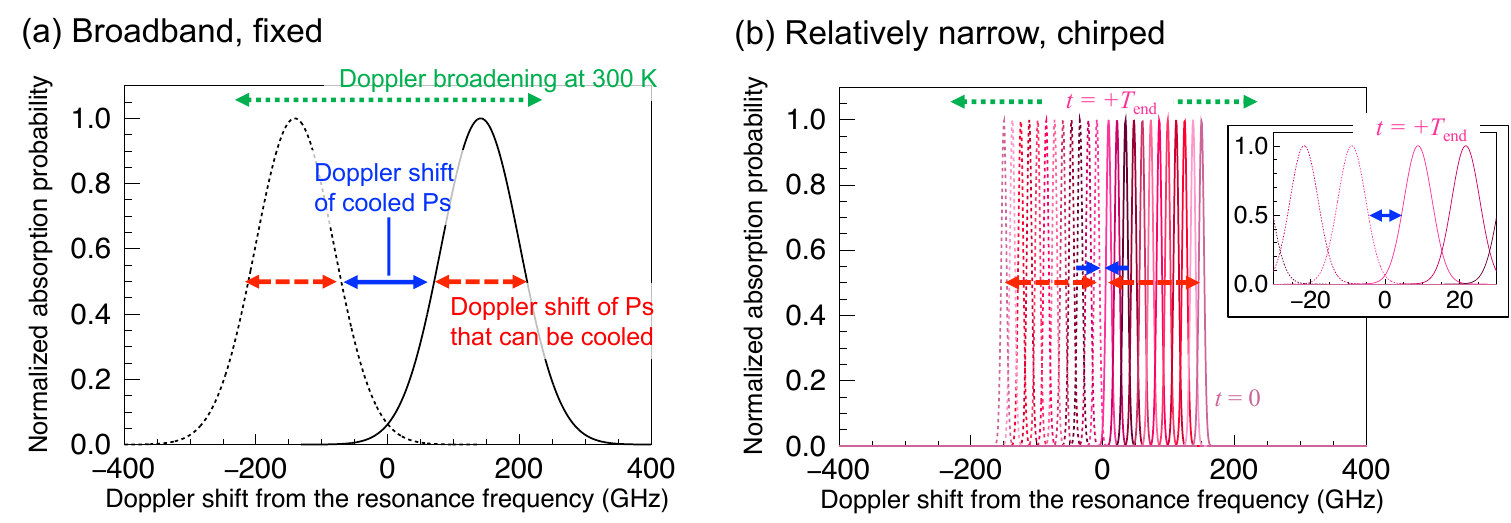}
  \caption{%
  Comparison of laser cooling schemes using a laser with a fixed spectrum and a chirped pulse train.
  It was assumed that the central frequency of the laser was red-shifted with respect to the 1S--2P resonance frequency and that the laser beam was irradiated in a counter-propagating configuration.
  The solid and dashed curves show the normalized absorption probability as a function of the Doppler shift of Ps from the 1S--2P resonance frequency for both beams.
  (a) Spectrum of the laser is fixed.
  The linewidth of the laser should be as broad as the Doppler broadening of Ps.
  FWHM of the Doppler broadening at \SI{300}{K} is indicated by dotted arrows to represent the typical broadening.
  The frequency detuning with this fixed broadband spectrum should be as large as the linewidth to avoid heating.
  Under these conditions, the Doppler broadening of the cooled Ps, indicated by the solid arrow, is comparable to the range of the Doppler shift of Ps that can be cooled (dashed arrow).
  (b) Situation in which a chirped pulse train is used.
  The shift is designed such that frequency detuning with respect to the 1S--2P transition frequency decreases with elapsed time, as shown by the spectra in different colors.
  A Ps atom that initially has a high velocity can be cooled by a pulse at an earlier time and then by the succeeding pulses, which can always be resonant with the decelerating Ps atom.
  It is possible to apply cooling to a wide range of initial velocity distribution if the laser frequency chirps upward with an appropriate rate.
  It is desirable for the rate to be equal to the change in resonance frequency of Ps per unit time, which is determined by the number of cooling cycles per unit time.
  If the linewidth of the pulse is relatively narrow, the frequency detuning at time $+T_{\mathrm{end}}$, which is the end of the cooling process, can be considerably smaller without accelerating Ps than that in the case of a fixed spectrum.
  The inset shows an enlarged view of the area around the origin of the horizontal axis.%
  }
  \label{fig:CompareCoolingScheme}
\end{figure*}
\section{Concept of the design}
\subsection{Requirements for efficient cooling}
Owing to the light mass and finite lifetime of Ps, a laser for Ps cooling is required to have appropriate structures in both the time and spectral domains.
Since the first proposal of Doppler cooling of Ps~\cite{liang_laser_1988}, some researchers have pointed out that the laser should be a pulsed laser with a wavelength of \SI{243}{nm}, have a long time duration on the order of \SI{100}{ns}, and a broadband spectrum to cover the Doppler width~($27\sqrt{T/\SI{1}{K}}\,\mathrm{GHz}$ at the full-width of half maximum~(FWHM) at the temperature $T$) of a hot cloud of Ps atoms~\cite{kumita_study_2002,shu_study_2016,zimmer_positronium_2021}.
The wavelength is resonant with the transition between the 1S and 2P states.
The 1S--2P transition is the most suitable for Doppler cooling because of its short spontaneous emission lifetime of approximately \SI{3.2}{ns} and its long lifetime for decay to gamma rays~\cite{alonso_positronium_2016}.
The recoil momentum for the \SI{243}{nm} photon is \SI{5.10}{\eVperc}.
The required time duration can be estimated from the typical initial momentum of Ps, the effective rate at which spontaneous emission occurs, and the recoil momentum.
For example, the most probable momentum of the cloud of Ps in the Maxwell--Boltzmann distribution at the temperature $T$ is $13 \sqrt{T/\SI{1}{K}}\,\mathrm{eV}c^{-1}$.
When the transition is strongly induced to saturation, the effective rate of the cooling cycle is $\frac{1}{2} \times \frac{1}{3.2\,\mathrm{ns}}$.
Therefore, the duration required to cool Ps to the cooling limit was calculated to be $16 \sqrt{T/\SI{1}{K}} \, \mathrm{ns}$.
Cooling of Ps from several hundreds of Kelvin can be accomplished in a time duration comparable to the lifetime of spin-triplet Ps.
This rapid cooling is possible because of the high recoil velocity, which originates from the light mass of Ps.
A pulsed laser with this long duration, rather than a continuous-wave oscillation, is preferable to achieve sufficient intensity to maximize the effective rate of spontaneous emissions with a typical laser beam size to cover the size of a gaseous cloud of Ps.

The requirement for the spectral properties discussed in several studies ~\cite{liang_laser_1988,kumita_study_2002,shu_study_2016,zimmer_positronium_2021} is that the laser should be broadband to cover the Doppler profile of the gaseous cloud of Ps.
Here, we point out that providing the pulsed laser with an appropriate time-wise frequency shift (i.e., a chirp) is useful for decreasing the temperature limit that can be obtained by laser cooling.
This scheme of laser cooling is called chirp cooling.
\Cref{fig:CompareCoolingScheme} shows a comparison between schemes using a laser with a fixed spectrum and with a frequency shift.
In the case of a fixed laser spectrum, the frequency detuning for Doppler cooling should be as large as the broad linewidth of the laser to suppress the photon absorption process, which leads to the acceleration of Ps. 
The laser with this large detuning is predominantly resonant with a faster Ps, which experiences a large Doppler shift in the range indicated by the dashed arrow in \cref{fig:CompareCoolingScheme}~(a).
In contrast, the laser is off-resonant with a slower Ps, whose Doppler shift range is in the region indicated by the solid arrow in \cref{fig:CompareCoolingScheme}~(a).
Therefore, the efficiency of cooling of the slower Ps becomes low.
This problem can be solved by adopting a laser with a relatively narrow spectral width when measured in a time window equal to the spontaneous emission lifetime of the 2P state, but whose center frequency rapidly shifts upward in time~(up-chirps).
The frequency of the laser should be shifted by $\Delta_{\mathrm{r}}=\frac{\hbar c k_{\mathrm{p}}^2}{2 \pi m_{\mathrm{Ps}}}$ for every duration equal to the time required to complete a single cooling cycle.
The cycle consists of the absorption of a laser photon, followed by spontaneous emission. 
Here, $\Delta_{\mathrm{r}}$ is the decrease in the Doppler shift after absorbing a photon with wavenumber $k_{\mathrm{p}}$ traveling in the opposite direction to Ps; $c$ is the speed of light, $m_{\mathrm{Ps}}$ is the mass of Ps, and $\hbar$ is the reduced Planck constant.
With $k_{\mathrm{p}}=2\pi/\SI{243}{nm}$, $\Delta_{\mathrm{r}}$ is \SI{6.2}{GHz}.
To cool Ps initially with a Doppler width of hundreds of GHz within several hundred nanoseconds, the chirp rate should be on the order of \SI{100}{\GHz\per\us}.
Regarding the linewidth in the short time, an excessively narrow linewidth limits the amount of Ps that can be cooled.
One of the causes of this poor cooling efficiency is the energy splitting in the $2^3\mathrm{P}_J$~($J=0,1,2$) states, which results in approximately \SI{10}{GHz} splitting for the transition frequencies from the $1^3\mathrm{S}_1$ state~\cite{alonso_positronium_2016}.
With a linewidth of approximately \SI{10}{GHz}, the laser can excite all possible transitions that are used for laser cooling.
This linewidth is also desirable because it covers the residual Doppler width that originates from the randomness of the recoil direction in the spontaneous emission.
The laser can resonate with arbitrary Ps recoiled by a spontaneously emitted photon in any direction and its resonant frequency is shifted by $\Delta_{\mathrm{r}}$ at the maximum.
Therefore, the amount of Ps that becomes off-resonant with the laser during the chirp cooling process can be decreased.
This linewidth contributes to an additional Doppler broadening of laser cooled Ps atoms, but it is a small broadening corresponding to temperatures approximately equivalent to the recoil limit, that is, $T_{\mathrm{r}}=\frac{(\hbar k_{\mathrm{p}})^2}{k_{\mathrm{B}} m_{\mathrm{Ps}}}=\SI{0.30}{K}$, where $k_{\mathrm{B}}$ is the Boltzmann constant.
This low temperature that is limited by the nature of laser cooling is achieved by combining the unprecedentedly rapid chirp rate and the optimal linewidth for each constituent pulse.

\begin{figure*}
  \centering
  \includegraphics[width=0.8\linewidth]{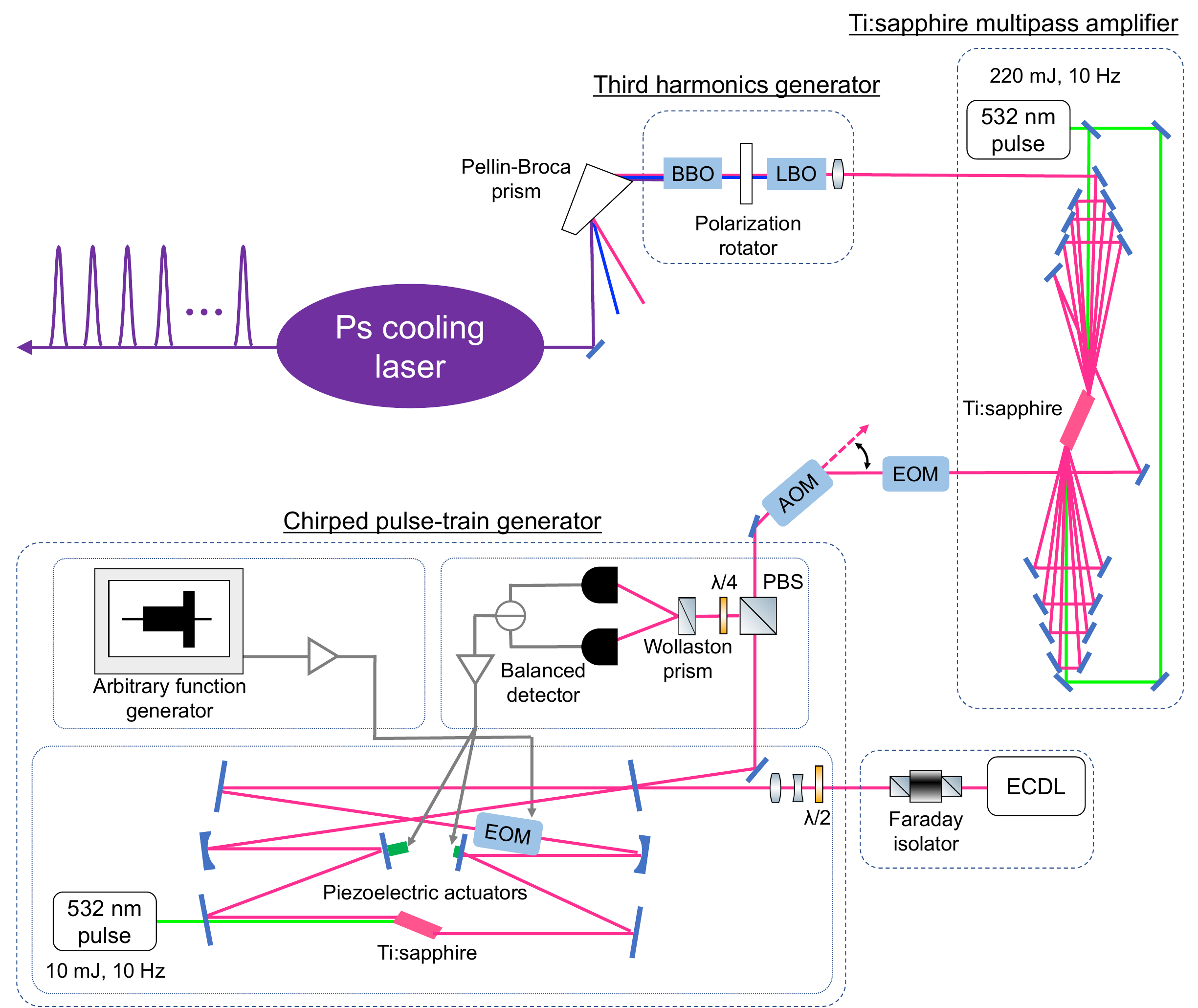}
  \caption{%
  Schematic of the laser system.
  The Ti:sapphire pulsed laser is injection-seeded by the external cavity diode laser~(ECDL), which is a \SI{729}{nm} continuous-wave~(CW) laser, so that the required \SI{243}{nm} laser can be obtained after the Ti:sapphire multipass amplifier and third harmonics generation~(THG) by the LBO and BBO nonlinear crystals.
  By driving the electro-optic modulator~(EOM) inside the laser cavity, the CPTG outputs a train of pulses.
  Each pulse has a broad linewidth and carrier-frequency chirp.
  The output of the CPTG is then diffracted by an acousto-optic modulator~(AOM) before being sent to the subsequent stages.
  A desired part of the CPTG output can be processed by controlling the timing of the pulsed drive of the AOM and the excitation of the multipass amplifier.
  This pulse chopping can control the duration of the pulse train, which was also used for time-resolved spectroscopy of the laser in this study.
  The output is also modulated by the second EOM to make the spectrum dense for the 1S--2P transition of Ps.%
  }
  \label{fig:LaserSchematics}
\end{figure*}
\subsection{Design of the laser}
A laser with optimized properties for the chirp cooling of Ps can be achieved by enhancing the spectral features of the CPTG, the details of which are described in~\cite{yamada_theoretical_2021}.
A schematic of the laser system is shown in \cref{fig:LaserSchematics}.
The injection-seeded pulsed laser with an optical phase modulator introduced inside the laser cavity is called the chirped pulse-train generator because it emits a train of optical pulses with a short duration, whose carrier frequencies are shifted by each pulse.
In the following section, we briefly review the essence of the working principle of the CPTG and introduce the necessary upgrades to fulfill the requirements for chirp cooling of Ps.

The requirements for the chirp cooling of Ps in the time and spectral domains were achieved by carefully selecting the parameters of any component of the laser.
For the time domain, a long cavity length~(\SI{3.8}{m}) and a relatively high reflectivity of the output coupler~(98\%) can be adopted for the long photon lifetime of the optical pulses that circulate inside the laser cavity.
The duration of the pulse train was more than \SI{600}{ns} at FWHM, which was sufficiently long.

Spectral broadening with chirp results from the phase modulation using an EOM inside the laser cavity.
The electric field of the light circulating inside the cavity $n$ times is multiplied by 
\begin{equation}
\begin{split}
  G^n & \exp (i n \beta \sin(\Omega_{\mathrm{m}} t)) \\
  &=\sum_{k=-\infty}^{\infty} G^n J_k(n\beta)\exp(ik\Omega_{\mathrm{m}}t),
\end{split}
  \label{eq:EOMModulationInsideCavity}
\end{equation}
where $G$ denotes the roundtrip gain of the cavity, $\beta$ is the modulation depth, $\Omega_{\mathrm{m}}$ is the modulation angular frequency, and $t$ is the elapsed time.
\Cref{eq:EOMModulationInsideCavity} shows that a sideband of the $k$-th order can be obtained if $G^{\frac{k}{\beta}} \gtrsim 1$ because the $k$-th order Bessel function of the first kind $J_k(x)$ has a significant value with $x \gtrsim k$.
Therefore, the amplitude of the largely modulated component is very different, depending on whether the value of $G$ is greater than unity.
In the operating sequence of the CPTG, either condition is fulfilled at different times as follows, which results in both spectral broadening and chirp.

The operation sequence of the CPTG begins with the accumulation of a continuous-wave, single-longitudinal-mode (SLM) seed laser inside the pulsed laser cavity.
Before the excitation of the laser medium, which was the Ti:sapphire crystal in this study, $G$ is less than unity, as determined by the total loss of the cavity.
The amplitude of the $k$th sideband $G^{\frac{k}{\beta}}$ is then a decreasing function of $n$.
In the steady state, the accumulated light inside the cavity consists of a superposition of the electric fields of light that have experienced different numbers of circulations from zero to infinity.
The superposed light has a broadband spectrum and is a train of pulses in the time domain.
The duration of each pulse was determined based on the broadness of the resulting spectrum.
The larger $G$ and $k$ lead to a broader spectrum.
The detailed numerical analysis that modeled the CPTG~\cite{yamada_theoretical_2021} showed that the linewidth is
\begin{equation}
  \frac{\beta \Omega_{\mathrm{m}} F \log 2}{2\pi}
  \label{eq:CPTGSinglePulseWidth}
\end{equation}
at FWHM~\cite{yamada_theoretical_2021}.
$F\simeq\pi/(1-G)$ denotes the laser cavity finesse.
The linewidth after the third harmonics generation (THG) will be broadened.
The width depends on the spectral phase in a single pulse.

A rapid chirp also appears from the EOM inside the cavity after the excitation of the laser medium by the nanosecond pulsed green laser.
The light at time $t$ from the excitation is a superposition of the electric fields that circulate in the cavity from zero to approximately $t/T_{\mathrm{trip}}$ times, counting from the excitation, where $T_{\mathrm{trip}}$ is the roundtrip time of the laser cavity.
When $G$ is larger than unity due to the excitation, the $t/T_{\mathrm{trip}}\beta$-th order sideband in the cavity becomes dominant.
Sidebands of each pulse also form a pulse with a similar linewidth.
As the order of the dominant sidebands increases over time, the pulse experiences a frequency shift of $\Omega_{\mathrm{m}} \beta$ for each time interval of $T_{\mathrm{trip}}$.
Therefore, the chirp rate of the angular frequency is $\Omega_{\mathrm{m}} \beta / T_{\mathrm{trip}}$, which triples after the THG to the desired wavelength of \SI{243}{nm}.

\begin{figure}
  \centering
  \includegraphics[width=8.6cm]{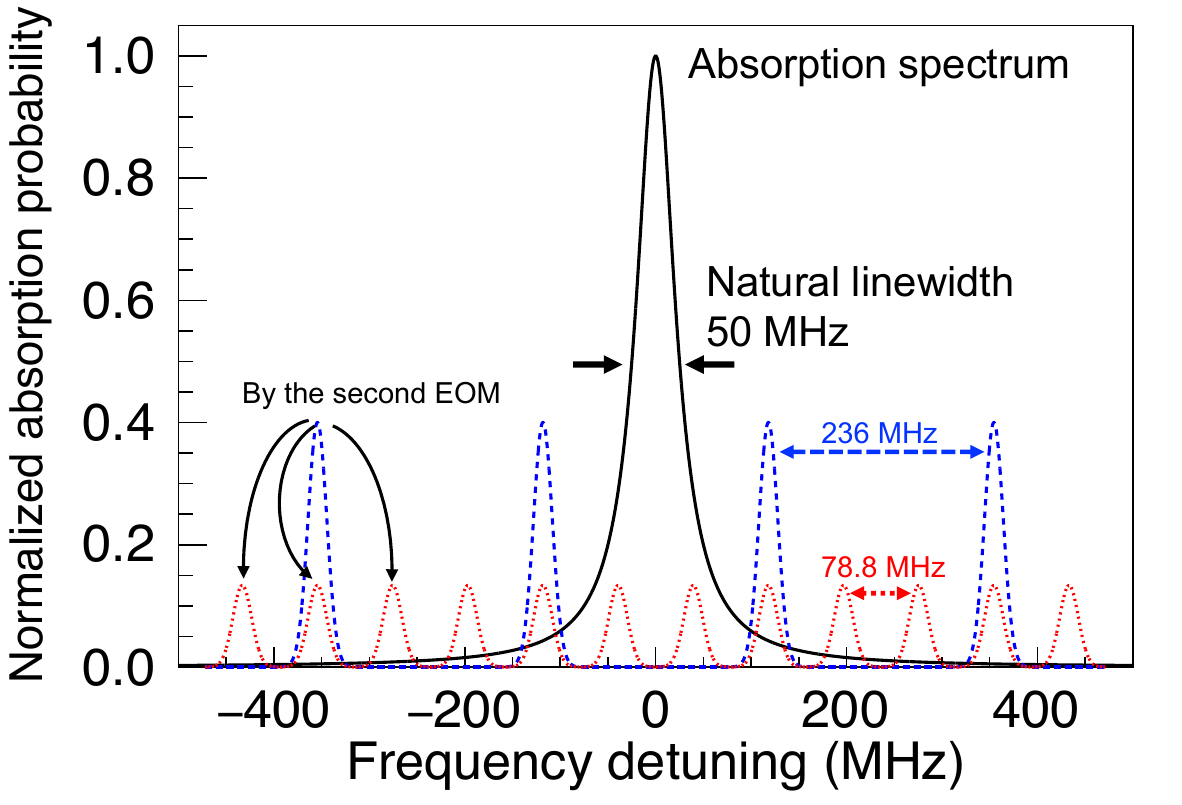}
  \caption{%
  Comparison between the natural linewidth of 1S--2P transition of Ps and the interval of the discrete spectrum of the laser.
  The absorption spectrum of Ps that is broadened by the natural linewidth of approximately \SI{50}{MHz} is shown as a centered solid curve.
  Because the developed CPTG has a discrete spectrum with an interval of \SI{236}{MHz} (dashed curves), there are many Ps atoms whose resonant frequencies are shifted by the Doppler effect and are off-resonant with any spectral component of the laser.
  Therefore, we generated sidebands with a modulation frequency around the natural linewidth to make the spectrum dense for the transition of Ps.
  We adopted $\SI{78.8}{MHz}$ as the modulation frequency for the second EOM, which is one-third of the interval of the unmodulated spectrum. 
  The spectra of the modulated laser are shown by the dotted curves.%
  }
  \label{fig:SpectrumShouldBeDense}
\end{figure}
The modulation angular frequency $\Omega_{\mathrm{m}}$ and depth $\beta$ were determined to obtain an optimized spectrum for Ps chirp cooling.
$\Omega_{\mathrm{m}}$ was set as an integer multiple of the longitudinal mode interval of the laser cavity such that the sidebands were also resonant with the cavity modes.
The interval was approximately \SI{78.8}{MHz} for the current laser system. 
The other parameters related to $\Omega_{\mathrm{m}}$ include the driving power and frequency of the second EOM, which was placed outside the laser cavity to modulate the output of the CPTG.
This EOM was used to make the laser spectrum sufficiently dense for Ps, which has a natural linewidth of approximately $\Gamma_{\mathrm{nat}}=\SI{50}{MHz}$ in the 1S--2P transition.
The CPTG output, whose spectrum is discrete with an interval of $\Omega_{\mathrm{m}}$, should be deeply phase modulated to fill the interval such that the frequency interval after the second EOM is less than or comparable to $\Gamma_{\mathrm{nat}}$.
A comparison between the natural linewidth and the interval of the discrete spectrum of the laser is presented in \cref{fig:SpectrumShouldBeDense}.
We adopted $\Omega_{\mathrm{m}}=2\pi\times\SI{236}{MHz}=2\pi\times\SI{78.8}{MHz}\times 3$ to balance the rapid chirp rate and difficulty of the second modulation.
Because a frequency shift occurs for each $T_{\mathrm{trip}}$ interval and it is longer than the duration of a cooling cycle when the transition is saturated, the required chirp rate with the current condition is $\Delta_{\mathrm{r}} / T_{\mathrm{trip}}=\SI{0.49}{GHz/ns}$, which is an order of magnitude higher than that achieved in~\cite{yamada_theoretical_2021}.
The modulation depth corresponding to this chirp rate is $\beta=\SI{8.8}{rad}$.
The linewidth of a single pulse that constitutes the output of the CPTG in the fundamental wave is \SI{22}{GHz}, which is broader than the target value.
Both requirements should be satisfied by dynamically controlling $\beta$ at a certain low value while operating the EOM before the laser excitation, and at the required level for a rapid chirp thereafter. 

\section{Pulsed driving of EOM for enhancing the spectral broadening}
We adopted a free-space resonantly enhanced EOM for the modulator inside the laser cavity to achieve the required chirp rate and single-pulse linewidth.
The free-space type was selected because of its low insertion loss, which is necessary for the high finesse of the laser cavity.
The EOM used in this work consists of a $\mathrm{KTiOPO_4}$~(KTP) EO crystal with dimensions of $\SI{3}{mm}\times\SI{3}{mm}\times\mathrm{L}\SI{6}{mm}$ and was manufactured to have a low insertion loss at the working wavelength of \SI{729}{nm} by using a high-quality crystal and anti-reflective coating.
The finesse of the laser cavity with the EOM was measured as approximately 97.
The enhancement in the modulation through resonance can be useful for achieving the required deep modulation because we drive the EOM at a single frequency of \SI{236}{MHz}.
The designed Q-factor of the resonant circuit was 98, which indicated that the power of the driving RF signal required to realize $\beta=\SI{8.8}{rad}$ was \SI{25}{W}.
Heat management for this high RF power was unnecessary because we drove the EOM for \SI{16}{\us} at a repetition rate of \SI{10}{Hz}, which corresponded to a duty ratio of \num{1.6e-4}.
The modulation bandwidth for the driving RF was \SI{2.4}{MHz}, which was limited by the Q-factor.
This bandwidth was wide enough to increase the RF power during the build-up time of the CPTG of several hundreds of nanoseconds.
The modulation depth can then be adjusted to achieve the required chirp rate and linewidth of a single pulse.

\begin{figure*}
  \centering
  \includegraphics[width=1.0\linewidth]{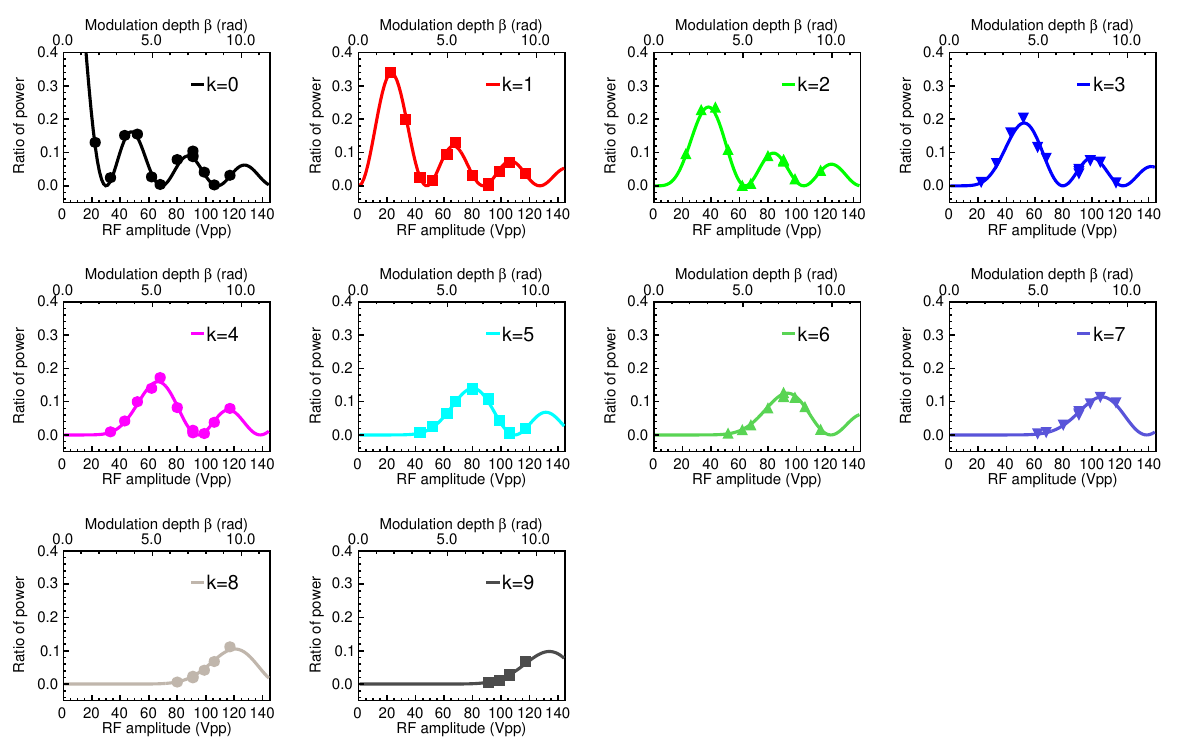}
  \caption{%
  Ratio of each sideband power to the incident power on the EOM as a function of the  amplitude of the driving RF.
  Peak-to-peak amplitudes are shown on the horizontal axis.
  The legend indicates the order of the resolved sideband.
  Sidebands up to the ninth order can be identified and their optical powers can be described well by the theoretical model of phase modulation~(\cref{eq:EOMModulationInsideCavity}), which is represented as solid curves.
  The conversion factor from the RF amplitude to the modulation depth $\beta$ was determined to be \SI{0.
  80}{rad/Vpp} to ensure that the measurements agreed well with the theoretical predictions.
  $\beta$ is indicated at the top of the horizontal axis.
  The maximum $\beta$ confirmed in this test was \SI{9.3}{rad}.%
  }
  \label{fig:SidebandsRatioVsRFAmplitude}
\end{figure*}

We tested whether the required modulation depth, which is too large for free-space EOMs with continuous driving, could be achieved in a pulsed operation.
The EOM was driven by a pulsed \SI{236}{MHz} RF signal, and the modulation depth was measured by analyzing the spectral broadening imposed on the SLM CW laser.
The modulated laser beam was injected into a scanning Fabry--P\'{e}rot interferometer~(\SI{1.5}{GHz} FSR, \verb|>|200 finesse), whose cavity length was swept such that its longitudinal modes were swept at a rate of \SI{4}{\MHz\per\s}.
Because the EOM was driven at a repetition rate of \SI{10}{Hz}, the optical frequency transmitted through the interferometer was shifted by \SI{0.4}{MHz} per driving pulse.
The transmittance through the interferometer was measured for \SI{40}{\micro s} using photodetectors~(\SI{350}{MHz} bandwidth), whose outputs were recorded at \SI{1.25}{GSa/s} with the EOM driven at \SI{10}{\micro s} after the beginning of the measurement.
The duration of driving the EOM was \SI{20}{\micro s}.
The optical spectrum was measured by sweeping the interferometer for \SI{1800}{s}, and the modulation depth was estimated from the distribution of the sideband power.
\Cref{fig:SidebandsRatioVsRFAmplitude} shows the ratio of each sideband power to the incident power on the EOM, as a function of the amplitude of the driving RF signal.
We were able to identify up to the ninth-order sideband, and the relationships between the sidebands were consistent with the theoretical model described by the Bessel functions in \cref{eq:EOMModulationInsideCavity}.
The conversion factor from the RF amplitude to $\beta$ that satisfactorily explains the measured ratios of each sideband power to the incident power was \SI{0.080}{rad/Vpp}.
The EOM worked properly up to the maximum driving amplitude, giving $\beta=\SI{9.3}{rad}$.
This is more than the depth required to achieve the optimal chirp rate.

To dynamically control the modulation depth to realize both the optimal linewidth of a single pulse and chirp rate, we adopted an arbitrary function generator (AFG) as the source of the driving RF signal.
The AFG outputs a sinusoidal wave at \SI{59}{MHz}, the amplitude of which is modulated by a square-wave envelope.
The amplitude was increased at the timing of the laser excitation of the CPTG.
The output frequency was then quadrupled by two serialized doublers, followed by a bandpass filter to allow frequency components around \SI{236}{MHz}.
The \SI{236}{MHz} pulse was amplified using a low-noise amplifier and a \SI{50}{W} power amplifier.
The typical peak-to-peak amplitudes of the resulting RF signals to obtain the required laser spectrum were \SI{30}{V} and \SI{141}{V}.
The latter amplitude was greater than \SI{110}{V}, which corresponded to $\beta=\SI{8.8}{rad}$ and provided the required chirp rate.
This was due to the slow rise time of the amplitude modulation of approximately \SI{400}{ns}, which was attributed to the response of the multipliers and filters. 
With this larger amplitude, we expected that the required $\beta$ for the chirp rate could be obtained immediately after the build-up of the CPTG.

It is noteworthy that a modulation that is too deep in the build-up period of the CPTG prevents the chirped pulse train from oscillating.
This was observed by driving the EOM with a constant amplitude.
By increasing $\beta$, we observed that pulses oscillated with a larger frequency shift from the seed laser, but in ranges greater than $\beta\simeq\SI{6}{rad}$ range, those pulses became weak and disappeared, and other pulses appeared dominantly at the seed frequency.
The frequency shift of the newly appearing pulses increased again when $\beta$ was further increased.
The smaller frequency shift from the seed laser indicates that the newly appearing pulses experienced fewer circulations in the cavity.
This can be explained by pulses whose frequency is shifted excessively from that of the seed laser, becoming off-resonant with the cavity.
In this case, the newly injected seed laser after the excitation becomes dominant in consuming the gain of the laser medium via the stimulated emissions.
The off-resonance of the largely shifted light originates from the unequally spaced longitudinal modes of the laser cavity, which are caused by the dispersive elements.
The longitudinal mode interval becomes detuned from the modulation frequency.
Sidebands generated in this region cannot interfere constructively with the electric fields that experience different numbers of circulations.
The EO crystal dominantly contributes to the dispersion by as much as ~ \SI{1e4}{fs^2} in the current setup.
This dispersion can limit the linewidth of a single pulse; however, the maximum linewidth available by limiting $\beta\simeq\SI{6}{rad}$ can be estimated from \cref{eq:CPTGSinglePulseWidth} as \SI{15}{GHz}, even in the fundamental wave.
This width is broader than \SI{10}{GHz}, which is the required value for the chirp cooling of Ps.
The described off-resonance problem can be neglected if a significant increase in $\beta$ occurs after excitation of the CPTG.
The power of a sufficiently built-up pulse train in the laser cavity can be very high when compared with that of the seed laser injected after the excitation.
Therefore, the gain of the laser medium is consumed by the original pulse train.
This holds true for our laser system, where we apply the amplitude modulated RF pulse.
Thus, the maximum depth of modulation can be applied to achieve the required chirp rate.

\section{Measurements of spectral and timing characteristics}
\begin{figure}
  \centering
  \includegraphics[width=8.6cm]{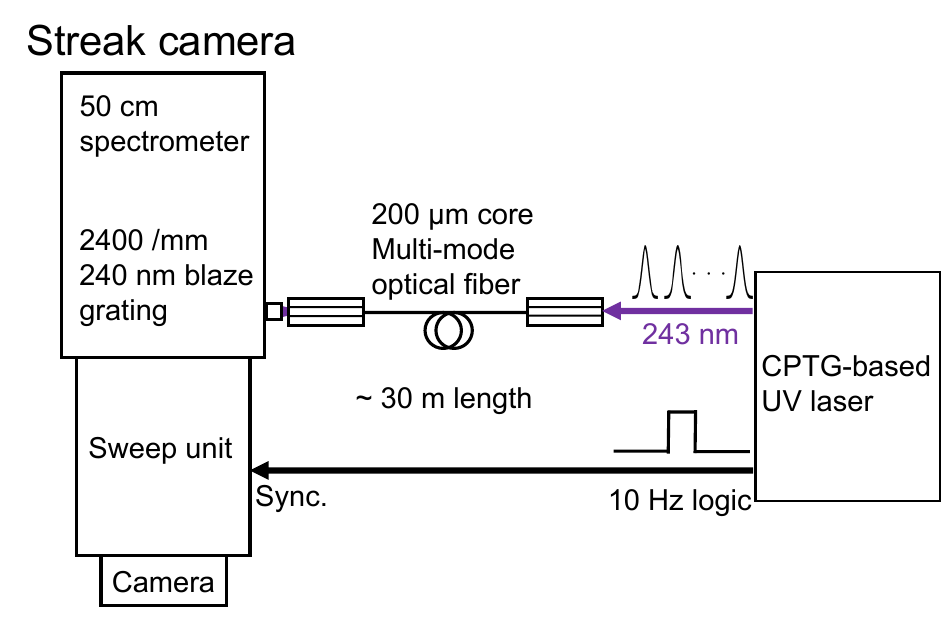}
  \caption{%
  Schematic setup for measuring the time-resolved spectrum using a streak camera.%
  }
  \label{fig:StreakSetup}
\end{figure}
\begin{figure*}
  \centering
  \includegraphics[width=0.8\linewidth]{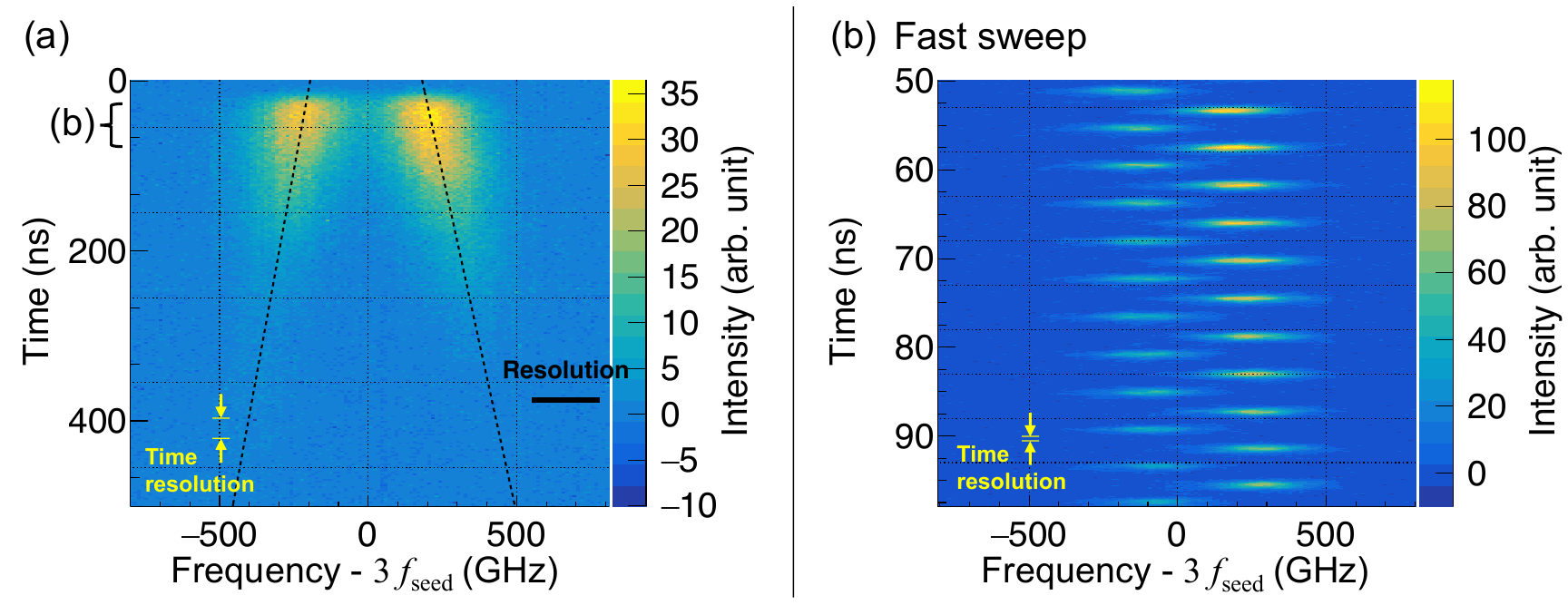}
  \caption{%
  Time-resolved spectra of the Ps cooling laser.
  (a) Spectrum acquired by setting the sweep time to observe all pulses in the train with duration of hundreds of nanoseconds.
  Two components with frequencies higher and lower than those of the seed laser were observed.
  Either component chirps upward ~(up-chirps) or downward~(down-chirps) so that the frequency shifts from the seed frequency become larger.
  The chirp rates were estimated to be \SI{6.e2}{\GHz \per \us} and \SI{-5.e2}{\GHz \per \us} by fitting the linear functions to the spectral data.
  The fitted functions are superimposed on the spectrum (dashed lines).
  (b) Time-resolved spectrum at the beginning of the pulse train,  acquired by fast sweeping to achieve better time resolution.
  The timing of the pulse trains of the two components are shifted by $\frac{1}{2}\frac{2\pi}{\Omega_{\mathrm{m}}}$, as predicted by the theoretical model~\cite{yamada_theoretical_2021}.
  They were well synchronized with the timing of the pulsed drive of the EOM.
  Both (a) and (b) are averaged over 100 shots.
  The origin of the horizontal axis corresponded to the optical frequency of the third harmonic of the seed laser denoted as $f_{\mathrm{seed}}$.
  The time resolutions were (a) \SI{23}{ns} at FWHM and (b) \SI{0.7}{ns} at FWHM.
  The frequency resolution was \SI{2.3e2}{GHz} at FWHM in both spectra.%
  }
  \label{fig:TimeResolvedSpectrum}
\end{figure*}
We operated the CPTG under the pulsed drive of the EOM, and measured the time-resolved spectrum of the pulse train by using a streak camera.
Because the spectral shift was enhanced by an order of magnitude compared to the system in~\cite{yamada_theoretical_2021}, direct observation of the time evolution of the laser spectrum became possible with the typical spectral resolution of the streak camera.
\Cref{fig:StreakSetup} shows the experimental setup of the measurement.
The laser was transmitted via a multimode optical fiber with a \SI{200}{\mu m} core. 
The streak camera was operated in the single-sweep mode, which was triggered by a \SI{10}{Hz} timing signal synchronized with the laser operation. Both the laser and trigger signals were transmitted over a distance of approximately \SI{30}{m}. The laser light was dispersed in a \SI{50}{cm} spectrometer with a grating with a groove density of \SI{2400}{\per \mm} and \SI{240}{nm} blaze wavelength.

\Cref{fig:TimeResolvedSpectrum} shows the typical time-resolved spectrum of the third harmonics of the CPTG, which is at \SI{243}{nm}.
In this measurement, the applied level of the RF peak-to-peak amplitude before the excitation of the CPTG was \SI{14}{V}.
In \cref{fig:TimeResolvedSpectrum}~(a), we can observe two components with time durations of hundreds of nanoseconds, whose frequencies are individually higher and lower than the seed frequency.
It is also clearly observed that both components shift away from the seed frequency.
By fitting a linear chirp function to the time-resolved spectra, the chirp rates were estimated to be \SI{6.2e2}{\GHz \per \us} and \SI{-5.2e2}{\GHz \per \us}.
The results demonstrate that a chirp rate that is more rapid than required can be obtained using the upgraded pulsed phase modulation.
\Cref{fig:TimeResolvedSpectrum}~(b) shows the spectrum acquired by the short sweep time to achieve a time resolution sufficient for resolving each pulse in the train.
Each chirped component formed a pulse train with a repetition rate same as the EOM modulation frequency $f_{\mathrm{m}}=\SI{236}{MHz}=1/\SI{4.23}{ns}$, not the longitudinal mode interval of the laser cavity.
With a tripled modulation frequency when compared with the mode interval, six pulses from both components circulate simultaneously in the laser cavity.
This is one of the different features when compared with the CPTG introduced in a previous study~\cite{yamada_theoretical_2021}, where the modulation frequency coincided with the mode interval of the cavity, and two pulses circulated in the cavity.
The time shift between two trains with opposite chirps was also directly observed as $\frac{1}{2f_{\mathrm{m}}}$, as predicted by the theoretical model~\cite{yamada_theoretical_2021}.

\begin{figure}
  \centering
  \includegraphics[width=7cm]{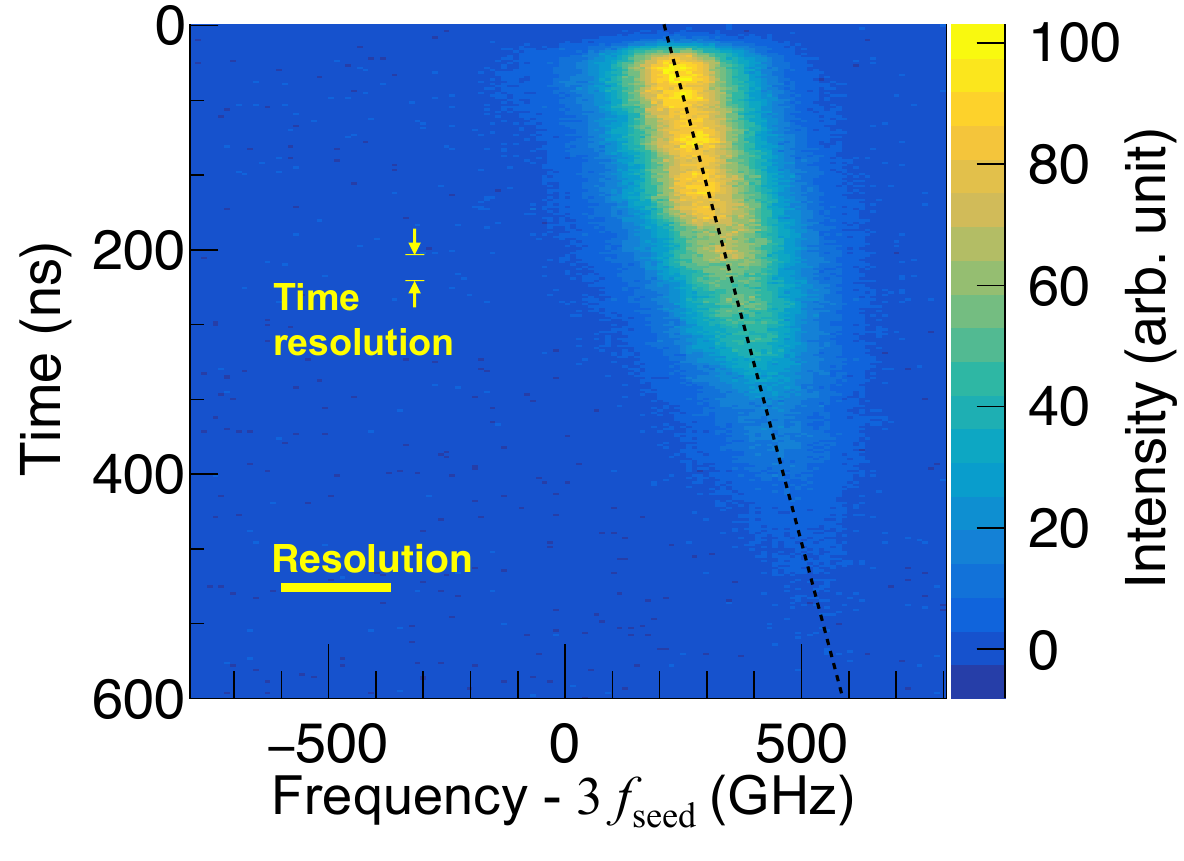}
  \caption{%
  Time-resolved spectrum with the up-chirp component optimized by adjusting the phase-matching angle.
  Because the phase-matching bands of both LBO and BBO crystals were centered in the frequency range of the up-chirp component, the THG efficiency for the chirping pulse train was maintained at a high value for a longer duration than that in \cref{fig:TimeResolvedSpectrum}~(a).
  The estimated chirp rate was \SI{6.
  e2}{\GHz \per \us}, which was obtained by fitting the superposed linear model.
  The THG efficiency of the down-chirp component was so low that it could not be observed.
  The time resolution was \SI{23}{ns} at FWHM, and the frequency resolution was \SI{2.3e2}{GHz} at FWHM.%
  }
  \label{fig:TimeResolvedSpectrumUpchirpOptimized}
\end{figure}
Because the up-chirp component is required for chirp cooling of Ps, we optimized the conversion efficiency of the THG for the up-chirp component by adjusting the phase-matching angles of the LBO~(\SI{10}{mm} length) and BBO~(\SI{4}{mm} length) crystals.
\Cref{fig:TimeResolvedSpectrumUpchirpOptimized} shows a typical time-resolved spectrum of the optimized up-chirp component.
The duration of the up-chirp component increased to approximately \SI{300}{ns} because the phase-matching bands of both LBO and BBO crystals were centered in the frequency range of the up-chirp component sweeps.
This duration is sufficiently long for chirp cooling of Ps from the typical initial temperature.
Under this phase-matching condition, the down-chirp component weakened because its frequency was outside the phase-matching bandwidth.
The linear chirp rate for the up-chirp component was estimated to be \SI{6.3e2}{\GHz \per \us}.
This result is consistent with the expected rate under the pulsed modulation of the EOM and is sufficiently rapid for realizing the optimized chirp cooling of Ps.
The optimum chirp rate of \SI{4.9e2}{\GHz\per\us} can also be achieved by adjusting the amplitude of the driving RF signal after the excitation of the CPTG to the lower value or by finely adjusting the timing of its amplitude modulation.

\begin{figure}
  \centering
  \includegraphics[width=8.6cm]{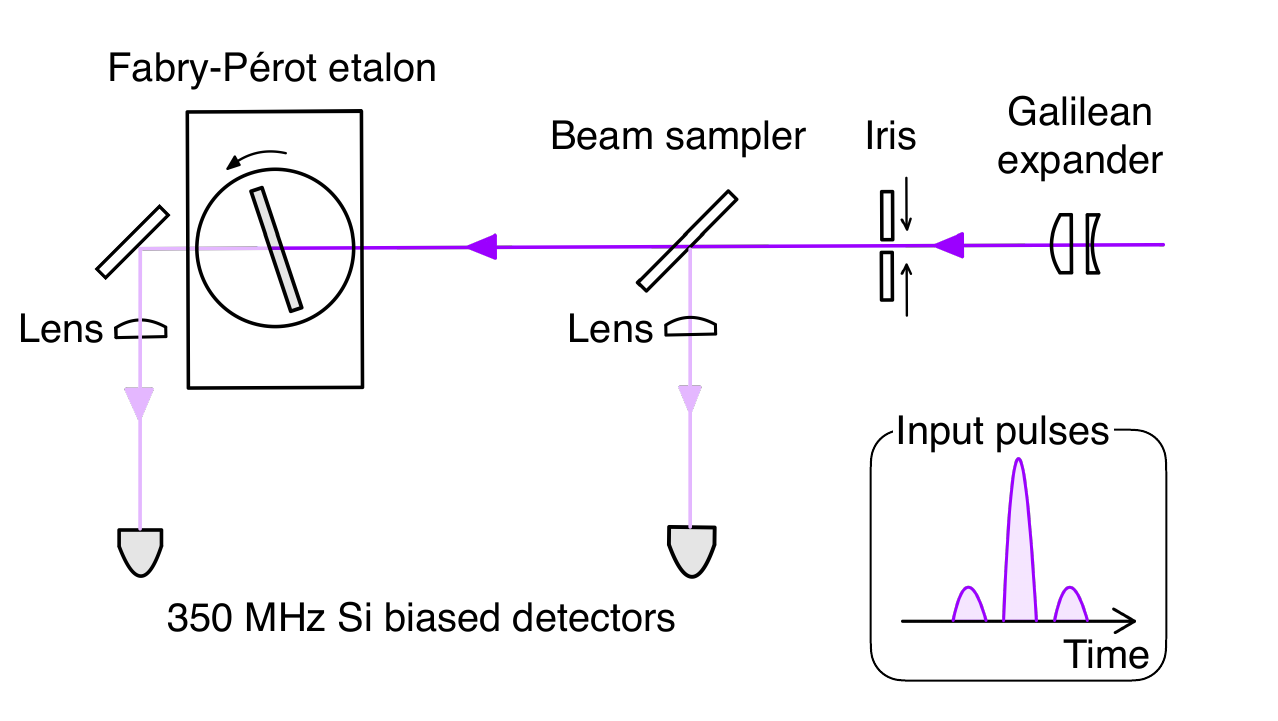}
  \caption{%
  Schematic setup for measuring the spectrum of a single pulse.
  The transmittance of each UV pulse was measured by scanning the etalon angle.
  To decrease the pile-up effect due to the limited bandwidth of the detectors used in our setup, the output of the CPTG was chopped to reduce the number of pulses to be detected.%
  }
  \label{fig:SinglePulseMeasumentSetup}
\end{figure}

\begin{figure}
  \centering
  \includegraphics[width=8.6cm]{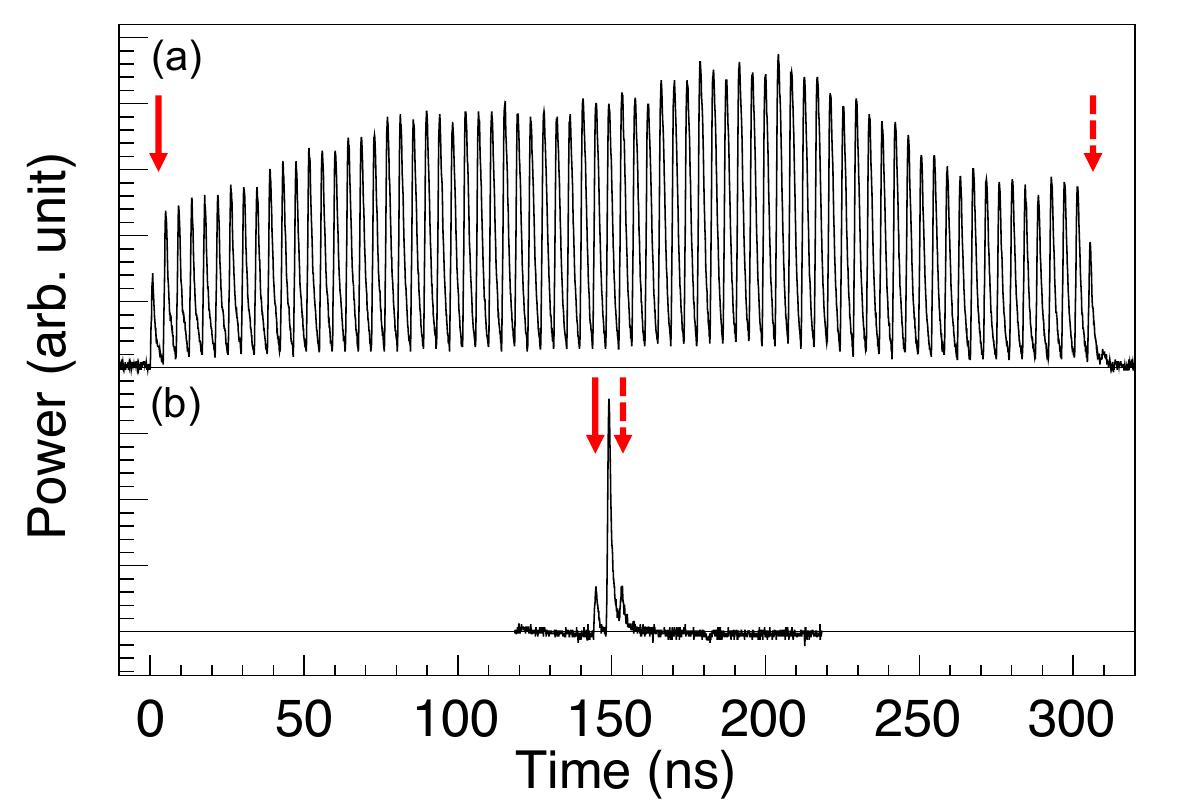}
  \caption{%
  Time-resolved optical power of the chopped pulse train at \SI{243}{nm}.
  The turn-on timings of the AOM and excitation timing of the multipass amplifier are indicated by solid arrows, and the turn-off timings of the AOM are indicated by dashed arrows.
  The duration of the pulse train can be controlled to (a) \SI{300}{ns} for the cooling of Ps and (b) approximately \SI{12}{ns} to measure the linewidth of a single pulse.
  Three pulses are observed in (b).
  The optical power of the first pulse was decreased by pulse chopping to reduce the pile-up effect on the second pulse, whose transmission spectrum was evaluated using an etalon.%
  }
  \label{fig:SinglePulsePower}
\end{figure}
Regarding the linewidth of each pulse, which is another important specification for the chirp cooling of Ps, we adopted another measurement method because the spectral resolution of the streak camera was not high enough to resolve the expected linewidth.
We used a Fabry--P\'{e}rot etalon as an optical narrow bandpass filter at \SI{243}{nm}, whose transmission frequency depends on the incident angle of light.
The optical power spectrum was obtained from the transmittance through the etalon by scanning the incident angle.
To resolve the spectrum of a single pulse, it is necessary to isolate the single optical pulse from the pulse train.
The setup for the measurement is shown in \cref{fig:SinglePulseMeasumentSetup}.
The incident and transmitted powers were measured using \SI{350}{MHz} bandwidth Si biased detectors.
Because the bandwidth was not wide enough to completely eliminate the falling tails of the preceding pulses, the pulse-chopping technique described in \cref{fig:LaserSchematics} using the AOM and multipass amplifier was adopted to reduce the number of pulses detected.
The typical time evolutions of the power measured at the input side are shown in \cref{fig:SinglePulsePower}.
For the measurements reported hereinafter, the applied level of the RF peak-to-peak amplitude before the excitation of the CPTG was \SI{30}{V}.
The timing of the operation of the AOM and the excitation of the multipass amplifier can control the duration of the pulse train. 
To measure the linewidth of a single pulse, the shortest duration of the pulse train, shown in \cref{fig:SinglePulsePower}~(b), was adopted.
The chopping bandwidth was not wide enough to extract a single pulse in the train, which can have a duration of \SI{300}{ns} for cooling Ps, as shown in \cref{fig:SinglePulsePower}~(a).
However, the power of the first pulse in \cref{fig:SinglePulsePower}~(b) decreased significantly such that the pile-up effect on the pulse of interest was negligible.

\begin{figure}
  \centering
  \includegraphics[width=8.6cm]{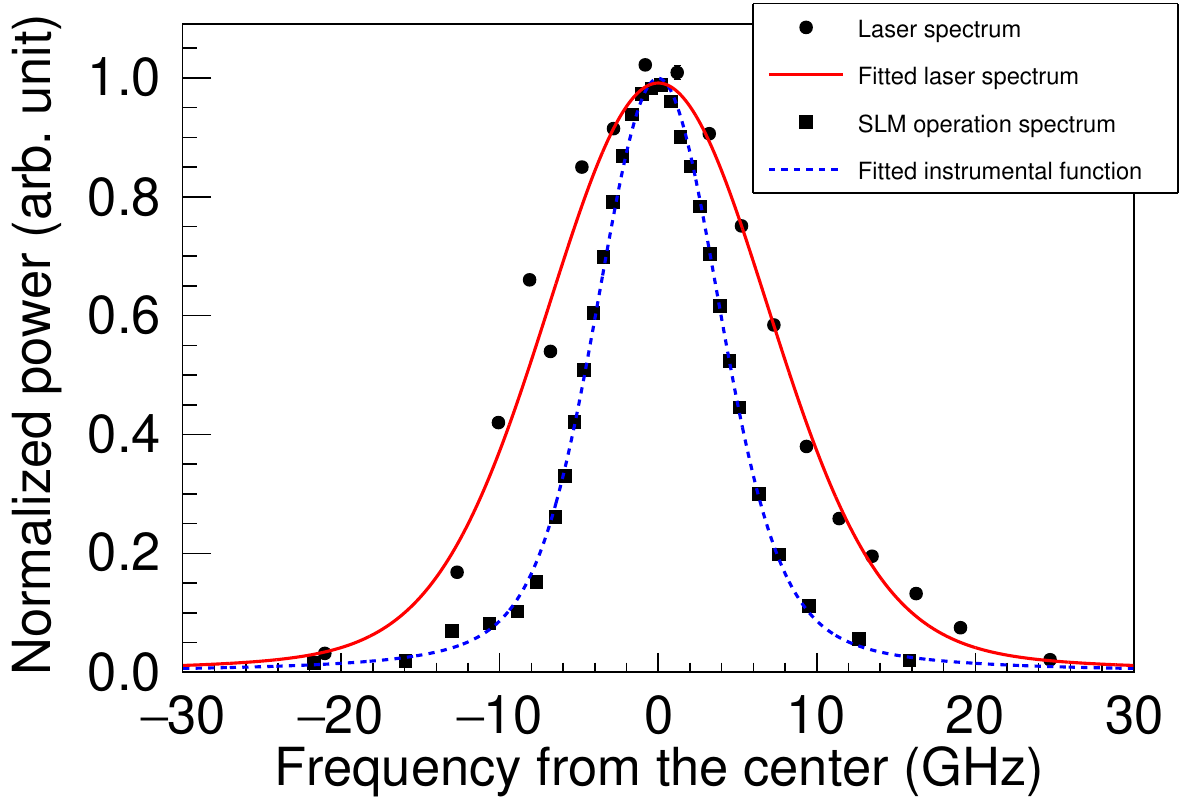}
  \caption{%
  Spectrum of a single pulse at approximately \SI{150}{ns} in the pulse train and instrumental functions.
  The instrumental function was evaluated using the SLM operation of the CPTG with a narrow linewidth.
  The measured spectrum under the SLM operation is indicated by filled squares.
  FWHM of the function was estimated to be \SI{10}{GHz} by curve fitting with a Voigt function, as indicated by the dashed curve.
  The measured spectrum of a single pulse is shown by the filled circles.
  The linewidth of a single pulse was estimated to be \SI{8.9}{GHz} at FWHM by subtracting the effect of the instrumental function.
  The subtraction was performed by the curve fitting of another Voigt function, assuming that the laser spectrum had a Gaussian lineshape and was convolved with the instrumental function.
  The fitted function is indicated by the solid curve. The spectral peaks are intentionally shifted to the origin of the horizontal axis.%
  }
  \label{fig:SinglePulseSpectrum}
\end{figure}
\Cref{fig:SinglePulseSpectrum} shows the measured spectrum of a single pulse at around \SI{150}{ns} in \cref{fig:SinglePulsePower}~(a).
We measured the instrumental function by measuring the transmission spectrum of the etalon using a SLM pulsed laser, which had a sufficiently narrow linewidth.
The measurement was performed by varying the operating optical frequency.
The SLM pulsed laser can be obtained from the CPTG without driving the EOM inside the laser cavity.
The frequency after THG was estimated by tripling the measured frequency of the seed laser for the CPTG using a wavemeter with an accuracy of \SI{60}{MHz}.
By subtracting the spectral resolution, the linewidth of a single pulse constituting the cooling laser was estimated to be \SI{8.9}{GHz} at FWHM, which is close to the designed value for optimal chirp cooling given by \cref{eq:CPTGSinglePulseWidth} and the applied $\beta=\SI{2.4}{rad}$ before the excitation of the CPTG.
We also measured the linewidth of each pulse at other times from \SI{0}{ns} to \SI{300}{ns} using a \SI{50}{ns} step, and the results were similar, ranging from \SI{8.2}{GHz} to \SI{9.8}{GHz} at FWHM.
This variation could originate from several factors such as the mismatch between the longitudinal mode interval of the CPTG cavity and the EOM modulation frequency, the mismatch between the seed laser frequency and the longitudinal mode, or the dispersion inside the cavity.

\begin{figure}
  \centering
  \includegraphics[width=8.6cm]{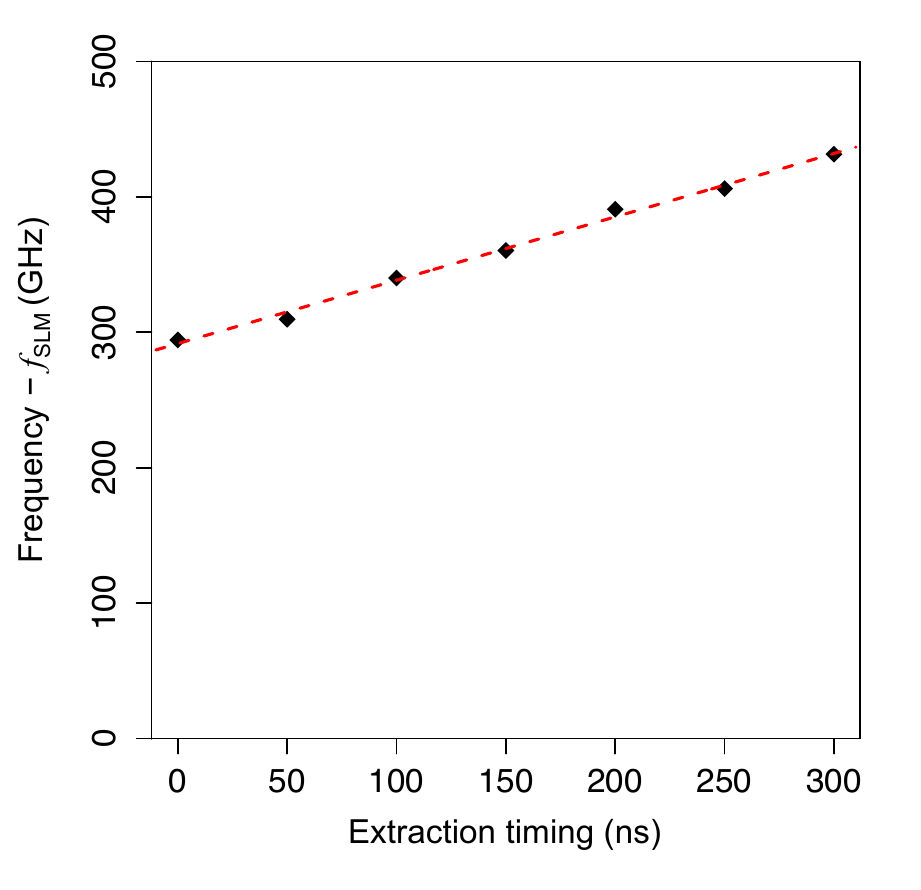}
  \caption{%
  Optical frequency of each pulse constituting a UV pulse train.
  $f_\mathrm{SLM}$ denotes the frequency of the SLM operation, which is the same as the third harmonic of the seeded frequency of the CPTG.
  The filled diamonds show the center frequencies of the extracted pulses.
  A linear chirp was observed at a rate of \SI{4.7e2}{\GHz \per \us}, which was estimated by fitting a linear function~(dashed line) to the data.%
  }
  \label{fig:FrequencyTimeEvolutionByWavemeter}
\end{figure}
Another application of this technique for extracting a few pulses from the pulse train at an arbitrary timing is the measurement of the chirp rate using a high-resolution wavemeter.
This measurement, which is complementary to that obtained by a streak camera, can be helpful, for example, in conducting laser cooling experiments at a positron beam facility where a streak camera is not available.
\Cref{fig:FrequencyTimeEvolutionByWavemeter} shows the frequency of the extracted pulse after taking the THG as a function of the extraction timing.
The frequency was measured using a UV-compatible wavemeter with an accuracy of \SI{\pm15}{GHz} and a resolution of \SI{40}{GHz} at \SI{243}{nm}.
We observed that the frequency of the pulse train starting from \SI{292}{GHz} shifted from the tripled seed frequency, and the pulses were linearly chirped at an estimated rate of \SI{4.7e2}{\GHz \per \us}.
An optimal chirp rate of \SI{4.9e2}{\GHz \per \us} was almost achieved; therefore, efficient chirp cooling was expected.
The difference from the result obtained using the streak camera originates from the difference in the modulation depth of the EOM and in the multipass amplifier in terms of the gain and amplification timing.
The multipass amplifier was adjusted to ensure that the power of the pulse train was as uniform as possible over time.

With the achieved chirp rate and linewidth of each pulse, we expect that the laser cooling of Ps is possible considering realistic experimental parameters such as the initial temperature and velocity measurement capability.
To estimate the area that can be irradiated with sufficient intensity by the laser to efficiently induce the 1S--2P transition, we derived the effective power per spectral component of the laser and considered how the laser beam can be large at an intensity comparable to the saturation level.
With the duration of the pulse train $\tau$ and linear chirp rate $R_{\mathrm{c}}$, the integrated linewidth is $\tau R_{\mathrm{c}}$.
The pulse energy per component of the laser spectrum with an interval of $f_{\mathrm{L}}$ is $\frac{U f_{\mathrm{L}}}{\tau R_{\mathrm{c}}}$, where $U$ is the energy of the pulse train.
Owing to the chirp, the effective duration of a single spectral component becomes shorter than the duration of the pulse train.
With the linewidth of a single pulse $\delta$, the effective duration can be estimated to be $\delta/R_{\mathrm{c}}$.
Therefore, the effective peak power per spectral component is $\frac{U f_{\mathrm{L}}}{\tau \delta}$.
The parameters of the laser developed in this study are $U=\SI{700}{\uJ}$, $\tau=\SI{300}{ns}$, $f_{\mathrm{L}}=\SI{78.8}{\MHz}$, and $\delta=\SI{8.9}{\GHz}$.
The calculated effective power was \SI{21}{\W} and the corresponding area of laser irradiation at the saturation intensity, which was \SI{0.45}{\W\per\square\cm}, was \SI{46}{\square\cm}.
This area is sufficiently wide to cover a flight distance of approximately \SI{2}{cm} for \SI{300}{ns} with the most probable velocity at \SI{300}{K}.
Another aspect to consider is the extent to which the velocity distribution of Ps can be narrowed by chirp cooling.
The frequency range that the laser sweeps due to the chirp is $\SI{4.7e2}{\GHz\per\us}\times\SI{300}{ns}=\SI{1.4e2}{GHz}$.
With two counter-propagating laser beams, Ps, whose Doppler shift for \SI{243}{nm} is within the doubled frequency range as wide as \SI{2.8e2}{GHz}, can possibly be cooled to the recoil limit temperature.
This range is more than half the FWHM~(\SI{4.6e2}{GHz}) of the Doppler broadening profile at \SI{300}{K}, and narrowing by chirp cooling can easily be detected by the well-established Doppler spectroscopy of Ps~\cite{cassidy_positronium_2010}.

\section{Conclusion}
We developed a pulsed laser at \SI{243}{nm}, which is optimal for the chirp cooling of positronium~(Ps).
Typically, the laser outputs a train of pulses whose carrier frequency was chirped at a rate of \SI{4.7e2}{\GHz\per\us} and the linewidth was \SI{8.9}{GHz}, both of which were directly confirmed by time-resolved spectroscopy measurements.
Based on the fundamental properties of the chirp-cooling scheme for Ps, these parameters are optimal for cooling as many Ps as possible to the recoil limit temperature for laser cooling, which is \SI{0.3}{K} for Ps.
The key aspect of the laser development was enhancing the spectral broadening of the chirped pulse-train generator, which we developed for Ps cooling, by an order of magnitude.
This enhancement was obtained by applying pulsed deep driving to the EOM.
Further, the developed laser has a suitable pulse energy for laser cooling of Ps prepared at typical temperatures.
The narrowing of the velocity distribution can be observed by the already established Doppler spectroscopy.
The chirp cooling using the developed laser will pave the way for precision spectroscopy and achieving the Bose--Einstein condensation of Ps.

\begin{acknowledgements}
We thank Dr. Yusuke Morita and Prof. Makoto Kuwata-Gonokami at The University of Tokyo for their assistance with measurements using the streak camera.
We are also grateful to Dr. Toshio Namba at The University of Tokyo for providing us with equipment to evaluate the linewidth of a single pulse.
This work was supported by JSPS KAKENHI Grant Number JP21K13862, and MEXT Quantum Leap Flagship Program, Grant Number JPMXS0118067246.
\end{acknowledgements}

\bibliography{ref.bib}

\end{document}